\documentclass[10pt,reqno]{article}

\usepackage{ulem}

\usepackage{amsmath,amssymb,amsthm}
\usepackage{esint}
\usepackage{bm} 
\usepackage{url}
\usepackage{subfigure}
\usepackage{graphicx,color}
\usepackage{algorithm}
\usepackage{algpseudocode}
\usepackage{algorithmicx}
\usepackage{hyperref}
\usepackage{float}
\usepackage{booktabs,multirow}
\usepackage{hhline}
\usepackage{setspace}

\usepackage{caption}
\usepackage{hyphsubst}
\usepackage{caption}

\usepackage{float}
\usepackage{hyperref}

\hypersetup{colorlinks,citecolor=blue,linkcolor=blue, urlcolor=blue}
\usepackage{pdflscape}

\newcommand{\RR}{\mathbb{R}}

\def\bx{{{\bf x}}}
\def\bpsi{{\boldsymbol{\psi}}}

\def\bz{{{\bf z}}}
\def\by{{{\bf y}}}

\DeclareMathAlphabet{\itbf}{OML}{cmm}{b}{it}

\newtheorem{thm}{Theorem}[section]

\newtheorem{prob}[thm]{Problem}

\newcommand{\email}[1]{\protect\href{mailto:#1}{#1}}

\newcommand{\pathfigures}{Figures/}
\graphicspath{{\pathfigures}}

\date{} 
\begin{document}
	
\title{Multipolar Acoustic Source Reconstruction from Sparse Far-Field Data using ALOHA\footnotemark[1]}
	\author{
		Yukun Guo\footnotemark[2]
		\and
 		Shujaat Khan\footnotemark[3]
 		\and
		Abdul Wahab\footnotemark[4] \footnotemark[5] 
            \and
		Xianchao Wang\footnotemark[2] 
	}
	\maketitle
	
	\renewcommand{\thefootnote}{\fnsymbol{footnote}}
	\footnotetext[1]{The work of Y. G and X. W was supported by the National Natural Science Foundation of China (NSFC) grant 11971133. The work of A.W. was supported by Nazarbayev University, Kazakhstan through Faculty Development Competitive Research Grant Program (FDCRGP) grant 1022021FD2914 and the Social Policy Grant.}
	\footnotetext[2]{School of Mathematics, Harbin Institute of Technology, Harbin, P. R. China (\email{xcwang90@gmail.com}; \email{ykguo@hit.edu.cn})}
 	\footnotetext[3]{Department of Computer Engineering, College of Computing and Mathematics, King Fahd University of Petroleum \& Minerals, Dhahran, 31261, KSA (shujaat.khan@kfupm.edu.sa)}
	\footnotetext[4]{Department of Mathematics, School of Sciences and Humanities, Nazarbayev University, 53, Kabanbay Batyr Avenue, 010000, Nur-Sultan, Kazakhstan (abdul.wahab@nu.edu.kz)}
	\footnotetext[5]{Corresponding author: A. Wahab at \email{abdul.wahab@nu.edu.kz}.}
	\renewcommand{\thefootnote}{\arabic{footnote}}
\begin{abstract} 
		The reconstruction of multipolar acoustic or electromagnetic sources from their far-field signature plays a crucial role in numerous applications. Most of the existing techniques require dense multi-frequency data at the Nyquist sampling rate. The availability of a sub-sampled grid contributes to the null space of the inverse source-to-data operator, which causes significant imaging artifacts. For this purpose, additional knowledge about the source or regularization is required. In this letter, we propose a novel two-stage strategy for multipolar source reconstruction from sub-sampled sparse data that takes advantage of the sparsity of the sources in the physical domain. The data at the Nyquist sampling rate is \textit{recovered} from sub-sampled data and then a conventional inversion algorithm is used to reconstruct sources. The data recovery problem is linked to a spectrum recovery problem for the signal with the \textit{finite rate of innovations} (FIR) that is solved using an \textit{annihilating filter-based structured Hankel matrix completion approach} (ALOHA). For an accurate reconstruction, a Fourier inversion algorithm is used. The suitability of the approach is supported by experiments.
	\end{abstract}
	
	\noindent 	\textbf{Keywords:} ALOHA; compressed sensing;  inverse source problem; multipolar source; sparse data imaging
	
\section{Introduction}

Inverse source problems (ISPs) have numerous applications, particularly in the fields of biomedical imaging \cite{Thio23,Beltrachini21,Ammari12}, non-destructive testing \cite{Takahashi22}, telecommunication \cite{Leone18}, and atmospheric sciences \cite{Badia02, AmmariN}. Several algorithms have been developed for solving ISPs. 

Leone, Maisto, and Pierri \cite{Leone18} employed ISPs in electromagnetic media to \textit{synthesize conformal antennas}. Beltrachini et al. \cite{Beltrachini21} and Thio et al. \cite{Thio23} have used ISPs to simulate neuron responses for intracranial recordings and for \textit{electroencephalography} (EEG). In a similar vein, ISPs have been used in \textit{magnetoencephalography} (MEG), for example, in \cite{Jerbi04}. An ISP for the reconstruction of temporally localized acoustic sources for \textit{photoacoustic tomography} has been dealt with in \cite{Ammari12}. Given their potential uses in security robots, cross-correlation-based passive imaging algorithms were suggested for locating correlated ambient noise sources in \cite{AmmariN}. A deep learning \textit{elastography} framework that uses sparse dynamic measurements was suggested in \cite{Yoo2018}.
On the technical side, an optimization framework was developed for reconstructing the minimal-energy sources in \cite{Devaney}. A finite element method was proposed in \cite{Beltrachini19} for the detection of multipolar sources with applications in EEG.  
\textit{Multiple Signal Classification} (MUSIC) algorithms for ISPs were discussed in \cite{Griesmaier017b}. 

The only measurements of the radiated waves available in the majority of real-world ISPs and imaging setups are multi-frequency discrete measurements. This stymies many mathematical methods. The existence of non-radiating sources raises concerns about the uniqueness of a solution. Mathematical discussions on the issue of multi-frequency discrete measurements may be found in \cite{Bleistein}. For these issues, several mathematical algorithms have been put forth. Potential workarounds include minimum energy solutions \cite{Devaney}, regularization \cite{Saman}, or use of \textit{\`{a} priori} knowledge \cite{Kusiak1}. By using measurements over a range of frequencies in an open interval, certain strategies overcome the non-uniqueness issue; see, e.g., \cite{Eller09, Bao15SNA, Alzaalig20,Wang, Griesmaier17}.

In this letter, we propose a numerical strategy for solving the ISPs of imaging multipolar acoustic sources from sparse multi-frequency far-field data. The suitability of multipolar sources for localization of \textit{epileptogenic} sources using EEG \cite{Beltrachini21, Beltrachini19}, neuron response modeling using \textit{stereo-EEG} (sEEG) \cite{Thio23}, \textit{cortical cavity} localization using MEG \cite{Jerbi04},  and \textit{confocal antenna synthesis} \cite{Leone18} serves as the impetus behind our work.

In the past, factorization \cite{Griesmaier17}, Prony's  \cite{Potts13}, MUSIC \cite{Griesmaier017b}, and compressed sensing \cite{Fannjiang10} algorithms have been used to identify multipolar acoustic sources using sparse far-field data. However, the advantage of the inherent sparsity of multipolar sources in the physical medium has never been taken. Here we take advantage of this sparsity to \textit{recover} an \textit{enriched} set of measurements at the Nyquist sampling rate and then adopt a Fourier inversion approach \cite{Wang}  to reconstruct sources. Based on the theory of the signal with the \textit{finite rate of innovations} (FRI) \cite{vetterli2002sampling}, we relate the problem of the recovery of measurements to the low rankness of a structured Hankel matrix and use an \textit{annihilating filter-based low-rank Hankel matrix completion approach} (ALOHA) \cite{jin2015annihilating, ye2016compressive}. 
 
In Section \ref{s:form}, we formulate the ISP. In Section \ref{s:ALOHA}, we discuss the proposed algorithm. Numerical results are provided in Section \ref{s:Num}. The conclusion is provided in Section \ref{s:conc}.
	
\section{Mathematical Formulation}\label{s:form} 

Assume that $S(\bx)$ is a frequency-independent multipolar source function defined by 
\begin{align}
S(\bx):=\sum_{j=1}^J\left(\lambda_j+\bpsi_j\cdot\nabla_\bx\right)\delta (\bx-\bz_j), \quad J\in\mathbb{N}_+,\label{S}
\end{align}
where $\delta$ is the Dirac mass and $\nabla_\bx$ is the distributional gradient with respect to $\bx\in\RR^d$, for $d=2,3$.  The points $\bz_1, \cdots, \bz_J\in\RR^d$ denote the locations of source components that are compactly embedded in the \textit{box} 
$\mathbb{A}:=(-{a}/{2}, {a}/{2})^d\subset\RR^d$, for $a\in\RR_+$, and 
constants $\lambda_j\in\RR$ and $\bpsi_j\in\RR_+^d$ are the corresponding intensities that satisfy conditions
\begin{align}
|\lambda_j|+|\bpsi_j|\neq 0 \,\text{ and }\, |\lambda_j\bpsi_j|=0 \quad\text{for}\quad j=1,\cdots, J.\label{cond1}
\end{align}

The radiation pattern, $u(\bx,k)$, of the source $S(\bx)$ satisfies
$ (\Delta+k^2)u(\bx,k)=S(\bx)$, for $\bx\in\RR^d$ and Sommerfeld's outgoing radiation condition (RC),
\begin{align}
\lim_{|\bx|\to\infty}|\bx|^{\frac{(d-1)}{2}}\left(\dfrac{\partial u}{\partial |\bx|}-\iota k u\right)(\bx,k)=0, \quad \iota:=\sqrt{-1}.\label{RC}
\end{align}
The parameter $k\in\RR_+$ is the so-called wavenumber and RC \eqref{RC} holds uniformly with respect to all directions 
$
\hat{\bx}:={\bx}/{|\bx|}\in\mathbb{S}^{d-1}:=\left\{ \bx\in\RR^d:\,\, |\bx|=1\right\},
$
where $\mathbb{S}^{d-1}$ represents the unit ball in $\mathbb{R}^d$. Here and throughout this article, a length-normalized quantity is marked by a superposed hat. 

The RC \eqref{RC} guarantees the existence of an analytic field $u_\infty:\mathbb{S}^{d-1}\times \RR_+\to\mathbb{C}$ that signifies the far-field radiation signature of $u$ and is defined by 
\begin{align*}
u(\bx,k)=\dfrac{e^{-\iota k|\bx|}}{|\bx|^{(d-1)/2}}
\left(u_\infty(\hat{\bx},k)+O\left(\dfrac{1}{|\bx|}\right)\right), \quad |\bx|\to\infty.
\end{align*}

To define the multi-frequency measurement grid, we introduce a parameter $\epsilon\in\RR_+$ such that $\epsilon\to 0^+$ and set 
$\mathbf{\ell}_0:= (\epsilon, 0)$ for $d=2$ or $\mathbf{\ell}_0:=(\epsilon,0,0)$ for $d=3$. We also introduce 
\begin{align}
\hat{\by}_{\mathbf{\ell}}&:=\hat{\mathbf{\ell}} \,\, \text{ for }\,\, \mathbf{\ell}\in\mathbb{Z}^d\setminus\{\mathbf{0}\}\quad\text{and}\quad  \hat{\by}_{\mathbf{0}}:=\hat{\mathbf{\ell}}_0, 
\label{xl}
\\
k_{\mathbf{\ell}}&:= \dfrac{2\pi|\ell|}{a}\,\,\text{ for }\,\,\mathbf{\ell}\in\mathbb{Z}^d\setminus\{\mathbf{0}\}\quad\text{and}\quad k_{\mathbf{0}}:=\dfrac{2\pi\epsilon}{a},
\label{kl}
\\
\nonumber
\mathbb{S}^{d-1}_{\rm dis}&:=\left\{\hat{\by}_{\mathbf{\ell}}\in\mathbb{S}^{d-1}\, \big|\quad \mathbf{\ell}\in\mathbb{Z}^d\right\}.
\end{align}

Let $\Omega_M:=\{\hat{\bx}_1,\cdots,\hat{\bx}_M\}\subset \mathbb{S}^{d-1}_{\rm dis}$
be the \textit{full set} of sampling points at the Nyquist sampling rate, i.e., the sampling points, $\hat{\bx}_m$, for $m=1,\cdots, M$, are chosen $k_{\rm min}/2$ distant apart from each other, where $k_{\rm min}:= \min\{k_1,\cdots, k_M\}$ with $k_m$ defined by \eqref{kl}, associated with the sampling point $\hat{\bx}_m$ defined by \eqref{xl}. 
We also define the \textit{sparse set} of measurement points, $\Omega_{M_R}$, as a (randomly chosen) subset of $\Omega_{M}$ for $R\ll M$, i.e.,   
$
\Omega_{M_R}:=\{\hat{\bx}_{M_r}\,|\quad r=1,\cdots, R\}\subset\Omega_M\subset \mathbb{S}^{d-1}_{\rm dis}.
$

 We consider the following inverse problems in this article. 

\begin{prob}[Full measurements ISP]\label{probF}
Recover the source $S(\bx)$ defined in \eqref{S} given the multi-frequency far-field data 
$$
\Big\{u_\infty(\hat{\bx}_{m}, k_{m})\,\Big| \quad\hat{\bx}_{m}\in\Omega_{M},\quad m=1,\cdots,M\Big\}.
$$
\end{prob}

\begin{prob}[Sparse measurements ISP]\label{probS}
Recover the source $S(\bx)$ defined in \eqref{S} given the multi-frequency far-field data 
$$
\Big\{u_\infty(\hat{\bx}_{M_r}, k_{M_r})\,\Big|\quad \hat{\bx}_{M_r}\in\Omega_{M_R}, \quad r=1,\cdots, R\Big\}.
$$
\end{prob}

\section{Hankel Matrix Completion Approach}\label{s:ALOHA}

The full measurement multi-frequency data ISP (Problem \ref{probF}) was solved in \cite{Wang} using a Fourier inversion technique. The source $S$ can be represented by the Fourier series     
\begin{align*}
S(\bx)=\sum_{\mathbf{\ell}\in\mathbb{Z}^d}\tilde{s}_{\mathbf{\ell}}\phi_{\mathbf{\ell}}(\bx)
\quad\text{with}\quad
\tilde{s}_{\mathbf{\ell}} =\dfrac{1}{a^d}\int_\mathbb{A} S(\bx)\overline{\phi_{\mathbf{\ell}}(\bx)}d\bx.
\end{align*}
Here, the bar indicates complex conjugate and  $\phi_{\mathbf{\ell}}(\bx):= e^{\iota \frac{2\pi}{a}\mathbf{\ell}\cdot \bx}$ for $ \mathbf{\ell}\in\mathbb{Z}^d$ are the basis functions.
The coefficients $\tilde{s}_{\mathbf{\ell}}$ can be calculated using the far-field measurements. To this end, the following result is proved in \cite[Theorem 2.1]{Wang}.
\begin{thm}\label{thm1}
Let $\hat{\bx}_{\mathbf{\ell}}$ and $k_{\mathbf{\ell}}$ be, respectively, defined by \eqref{xl} and \eqref{kl}, then the Fourier coefficients 
$\{\tilde{s}_{\mathbf{\ell}}\}_{\mathbf{\ell}\in\mathbb{Z}^d}$ of $S(\bx)$ can be determined uniquely by 
$\{u_\infty(\hat{\bx}_{\mathbf{\ell}}, k_{\mathbf{\ell}})\mid \mathbf{\ell}\in\mathbb{Z}^d\}$ and for any positive integer $N\to+\infty$,
\begin{align}
\tilde{s}_{\mathbf{\ell}}=&
-\dfrac{1}{a^d\gamma_d} u_\infty(\hat{\bx}_{\mathbf{\ell}}, k_{\mathbf{\ell}}), \qquad\mathbf{\ell}\in\mathbb{Z}^d\setminus\{\mathbf{0}\}, \label{eqsl}
\\
\tilde{s}_{\mathbf{\ell}_0}\approx&-\dfrac{\epsilon\pi}{a^d\sin(\epsilon\pi)\gamma_d} u_\infty(\hat{\bx}_{\mathbf{\ell}_0}, k_{\mathbf{\ell}_0})
\nonumber
\\
&+\sum_{1\leq |\mathbf{\ell}|_\infty\leq N} \tilde{s}_{\mathbf{\ell}} \int_\mathbb{A}\phi_{\mathbf{\ell}}(\by)\overline{\phi_{\mathbf{\ell}_0}(\by)}{\rm d}\by,\label{eqs0}
\end{align}
where $N$ is a truncation parameter to retain a finite number of terms in the inversion and
$\gamma_d:= {e^{\iota\pi/4}}/{\sqrt{8\pi k_{\mathbf{\ell}}}}$ for $d=2$ or $\gamma_d:={1}/{4\pi}$ for $d=3$.
\end{thm}

Given Theorem \ref{thm1}, $S$ can be reconstructed very accurately when \textit{enough} multi-frequency measurements are available at dense sampling points $\hat{\bx}\in\Omega_M$ at the Nyquist sampling rate, as all the Fourier coefficients can be calculated subject to measurement noise and an approximation of $\tilde{s}_{\mathbf{\ell}_0}$. However, the situation changes drastically when the recording grid is undersampled. A part of $S$ behaves like a non-radiated component since its radiation signature is not detected. The \textit{missing spectrum} corresponding to the \textit{unrecorded} sampling points and unused frequency profiles contributes to the null space of the inverse source-to-data operator due to under-sampling \cite{Yoo2018, ye2016compressive}. To this end, a traditional remedy is to use conventional or sparsity-promoting regularization methods to solve Problem \ref{probS} \cite{Fannjiang10}. Such regularization approaches are not very useful because the size of the speckle field introduced by the under-sampling is comparable to or even larger than the size of the sources (see Section \ref{s:Num}). In contrast, we suggest exploiting the sparsity of $S$ inside $\mathbb{A}\subset\RR^d$. In this section, we propose a two-stage numerical strategy to address Problem \ref{probS}. We assume a sampling at the Nyquist rate and that the given sparse measurements constitute a (random) part of the full grid measurements, while the rest are \textit{missing}. Subsequently, we link the missing spectral values to the low rank of a structured Hankel matrix and recover them using ALOHA. Once the full Nyquist grid measurements are recovered, we use Theorem \ref{thm1} to recover the source $S$.

Since $S$ is sparsely supported in the \textit{box} $\mathbb{A}$, it has a sparsely distributed Fourier spectrum in the low spatial frequency regions. Therefore, according to the sampling theory of a signal with a finite rate of innovations \cite{vetterli2002sampling},  there exists an annihilating filter, $\widetilde{\mathbf{h}}$, in the Fourier domain for the corresponding spectral vector $\tilde{\mathbf{s}}$ at the Nyquist sampling rate such that $(\widetilde{\mathbf{h}}\star\tilde{\mathbf{s}})_k:=\sum_{i=0}^{n}[\widetilde{\mathbf{h}}]_{i}[\tilde{\mathbf{s}}]_{k-i}=0$. Here, $[\mathbf{u}]_i$ represents the $i$th component of the vector $\mathbf{u}$,  $n+1$ is the length of  $\widetilde{\mathbf{h}}$, and $\star$ is the convolution operator. Consequently, the structured Hankel matrix (a sub-matrix of the corresponding convolution matrix), 
\begin{align*}
\mathcal{H}_p(\tilde{\mathbf{s}}):=
\begin{pmatrix}
\tilde{s}_1 & \tilde{s}_2 &\cdots & \tilde{s}_{p}
\\
\tilde{s}_2 & \tilde{s}_3 &\cdots & \tilde{s}_{p+1}
\\
\vdots & \vdots &\ddots & \vdots
\\
\tilde{s}_M & \tilde{s}_1 &\cdots & \tilde{s}_{p-1}
\end{pmatrix},
\end{align*}
associated to $\tilde{\mathbf{s}}\in\RR^M$ is rank-deficient if the \textit{matrix-pencil size} $p$ ($<M$) is chosen larger than the minimum filter size since 
$\mathcal{H}_p(\tilde{\mathbf{s}})\widetilde{\mathbf{h}}'=\mathbf{0}$,
where $\widetilde{\mathbf{h}}'$ is the flipped version of $\widetilde{\mathbf{h}}$. 
Specifically,  if the minimum length is $n+1$, then 
${\rm rank}\left(\mathcal{H}_p(\tilde{\mathbf{s}})\right)=n$ \cite[Theorem II.1]{ye2016compressive}.

Given \eqref{eqsl}-\eqref{eqs0} and measurements, $\{u_\infty(\hat{\bx}_{M_r}, k_{M_r})\,|\,\, \hat{\bx}_{M_r}\in\Omega_{M_R}\}$, the problem of recovering the missing spectrum $\{u_\infty(\hat{\bx}_{M}, k_{M})\,|\,\, \hat{\bx}_{M}\in\Omega_M\setminus\Omega_{M_R}\}$ for $\tilde{\mathbf{s}}$ can be formulated as the low-rank Hankel matrix completion problem,
\begin{align}
{\rm arg} \min_{\mathbf{g}\in\mathbf{C}^M} \|\mathcal{H}_p(\mathbf{g})\|_*  \text{ subject to }\,P_{\Omega_{M_R}}(\mathbf{g})=P_{\Omega_{M_R}}(\tilde{\mathbf{s}}),\label{opt-prob}
\end{align}
where $\mathcal{P}_{\Omega_{M_R}}$ is the projection on ${\Omega_{M_R}}$ and $\|\cdot\|_*$ is the matrix nuclear norm. 

Several algorithms can be used to solve the low-rank matrix completion problem \eqref{opt-prob} but we use ALOHA \cite{jin2015annihilating} in the proposed numerical scheme due to its robustness and performance guarantee. It is robust since it is a singular value decomposition (SVD)-free algorithm that uses a low-rank factorization model for initialization. Specifically, the optimization problem \eqref{opt-prob} is first converted to the constrained optimization problem 
\begin{align*}
& \min_{\left(\mathbf{U},\mathbf{V}\right)\in\mathbb{M}}\left(\|\mathbf{U}\|_{F}^2+\|\mathbf{V}\|_{F}^2\right), 
\\
&\text{subject to } \mathcal{H}_p(\mathbf{g})=\mathbf{U}\mathbf{V}^H
\text{ and } P_{\Omega_{M_R}}(\mathbf{g})=P_{\Omega_{M_R}}(\tilde{\mathbf{s}}),
\end{align*}
based on the observation that
\begin{align*}
&\|\mathcal{H}_p(\mathbf{g})\|_*= \min_{\left(\mathbf{U},\mathbf{V}\right)\in\mathbb{M}}\left(\|\mathbf{U}\|_{F}^2+\|\mathbf{V}\|_{F}^2\right), 
\\
&\text{where}\quad \mathbb{M}:=\left\{\left(\mathbf{U},\mathbf{V}\right)|\quad \mathcal{H}_p(\mathbf{g}) = \mathbf{U}\mathbf{V}^H\right\},
\end{align*}
and then solved using the \textit{alternating direction method of multipliers} (ADMM) \cite{ADMM}. Here the superposed $H$ represents the Hermitian transpose and $\|\cdot\|_F$ is the Frobenius norm.  Moreover, a unique minimizer to the matrix completion problem \eqref{opt-prob} is guaranteed with probability $1-1/M^2$ under a standard incoherence assumption on $\mathcal{H}_p(\mathbf{g})$ \cite{ye2016compressive}. Once a minimizer $\mathbf{g}^*$ is found, $S(x)$ can be reconstructed as
$$
S_{\rm recon}(\bx)=\sum_{m=1}^M [\mathbf{g}^*]_m\phi_{m}(\bx).
$$

\section{Numerical Experiments and Discussion}\label{s:Num}

\subsection{Experimental Setup}\label{ss:Setup}

For numerical simulations, we consider a two-dimensional setting. The source $S$ consists of two monopoles and two dipoles. The monopoles are located at positions $\bz_1:=(5,\,4)$ and $\bz_2:=(-4,\,-4)$ with magnitudes $\lambda_1:=9$ and $\lambda_2:=8$, respectively. The dipoles are located at positions $\bz_3:=(-4,\,5)$ and $\bz_4:=(4,\,-4)$ with intensities $\bpsi_3:=(1,\,-1)^T$ and $\bpsi_4:=(-1,\,1)^T$, respectively. Note that $\bpsi_1=\mathbf{0}=\bpsi_2$ and $\lambda_3=0=\lambda_4$ due to conditions \eqref{cond1}. 

Following definitions  \eqref{xl} and \eqref{kl}, the admissible sets of observation directions and wavenumbers are defined as
\begin{align*}
  \mathbb{X}_N:=&\big\{ \hat{\bx}_{\ell}= {\ell}/{|\ell|}\,\,\big|\,\, 1\leq|\ell|_{\infty}\leq N   \big\}\cup \left\{ (1,\,0) \right\},
\\
\mathbb{K}_N:=&\big\{ k_{\ell}={2\pi|\ell|}/{a}\,\,\big|\,\, 1\leq|\ell|_{\infty}\leq N   \big\}\cup \left\{ {2\pi\epsilon}/{a} \right\}.
\end{align*}
We set $a=12$, $\epsilon=10^{-3}$, and $N=20$. The synthetic far-field patterns are generated by numerical integration of the formula
\begin{equation*}
  u_\infty(\hat{\bx}_{\mathbf{\ell}}, k_{\ell} )=-\gamma_d \int_{\mathbb{A}} \
     S(\by) \, e^{-\iota k_{\ell} \hat{\bx}_{\ell}\cdot \by}\, d\by.
\end{equation*}

We perform source reconstruction in extremely noisy measurement conditions with an \textit{additive white Gaussian noise} (AWGN) having a \textit{signal-to-noise ratio} (SNR) of $3$dB. Three different sub-sampling rates, $5\%$, $10\%$, and $15\%$, were considered. The performance of the proposed algorithm was evaluated on: (1) visual perception in terms of the \textit{peak-signal-to-noise ratio} (PSNR), (2) structural similarity in terms of the \textit{structural similarity index measure} (SSIM), and (3) computational time in terms of walk-clock run-time.  Here,
\begin{align*}
{\rm SNR}&:= 10\log_{10} \left(\dfrac{\|v\|^2}{\|v-\tilde v\|^2}\right),
\\
{\rm PSNR}&:= 10 \log_{10}\left(\dfrac{v_{\max}^2}{\|v-\tilde v\|^2}\right),
\\
{\rm SSIM}&:=\dfrac{(2\mu_{v}\mu_{\tilde v}+c_{1})(2\sigma_{v,\tilde v} +c_{2})}{(\mu^{2}_{v}+\mu^{2}_{\tilde v}+c_{1})(\sigma^{2}_{v}+\sigma^{2}_{\tilde v} +c_{2})}
\end{align*}
where  $\mu_{v}$, $\mu_{\tilde v}$, $\sigma_{v}$, $\sigma_{\tilde v}$, and $\sigma_{v,\tilde v}$ are the means, standard deviations, and covariance for the signal $v$ and its noisy version $\tilde v$, calculated for a radius of $1.5$ units.  The constants $c_1$ and $c_2$ are set to $c_{1}=(0.01v_{max})^{2}$ and $c_{2}=(0.03v_{max})^{2}$.

Each reconstructed image obtained from sparse far-field data using ALOHA is compared with (1) \textit{original} image (with fully-sampled data at Nyquist rate), (2) \textit{input} image (with sub-sampled data without any regularization), and (3) \textit{L1-CS} image (with sub-sampled data using \textit{L1-compressed sensing}-based sparse regularization) \cite{8187544}.

For signal recovery, the number of filters in ALOHA was set to $15$, but all other default settings were retained. For the L1-CS, the all-platforms (v1.22) redistributable CVX package \cite{cvx} for specifying and solving convex programs was used. Since the visualization of monopole sources is somewhat limited in the input and L1-CS images, the scales of the individual images were adjusted. For computational time estimation, ALOHA and L1-CS were implemented on the MATLAB cloud platform running on an Intel(R) Platinum 8375C CPU @ $2.90$ GHz with a clock speed of $3464.412$ MHz and $128$GB RAM.

\subsection{Numerical Results}\label{ss:Results}

\begin{figure}[tbh!]
\begin{center}
\subfigure[Experiment with $5\%$ measurements]{\includegraphics[width = 1\columnwidth]{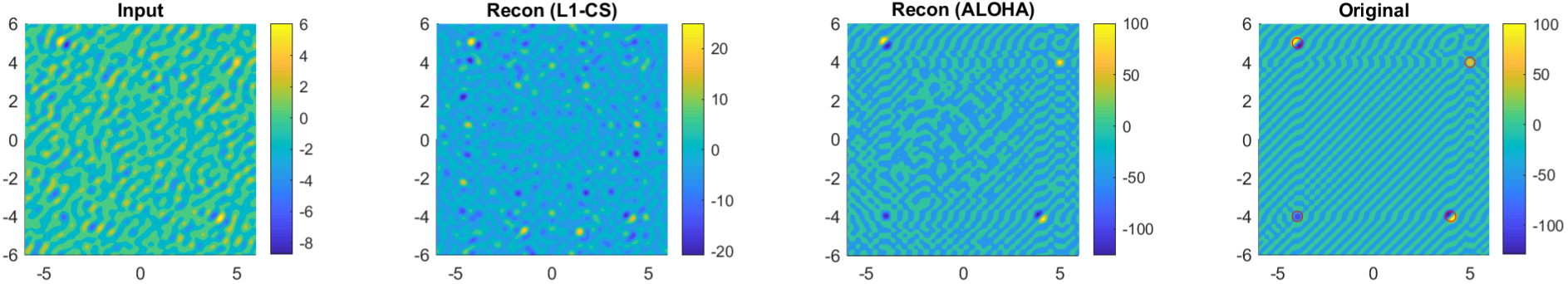}\label{fig2a}}
\\
\subfigure[Experiment with $10\%$ measurements]{\includegraphics[width = 1\columnwidth]{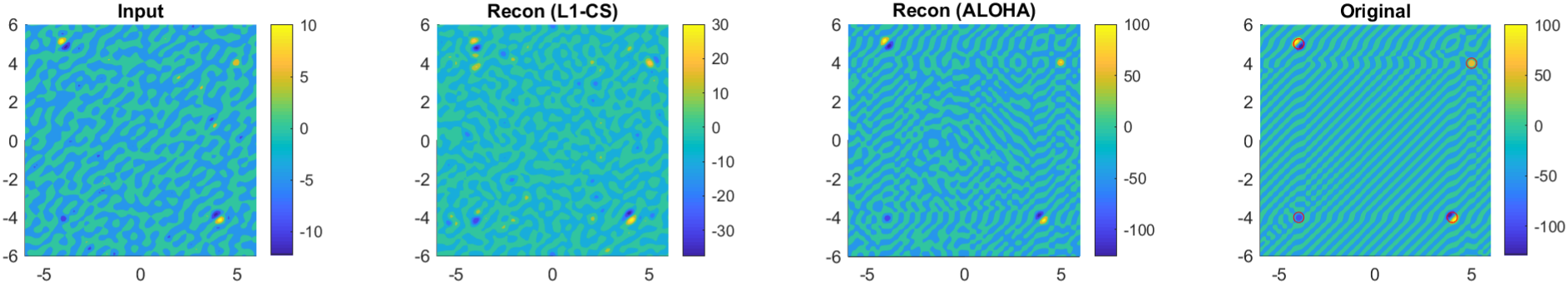}\label{fig2b}}
\\
\subfigure[Experiment with $15\%$ measurements]{\includegraphics[width = 1\columnwidth]{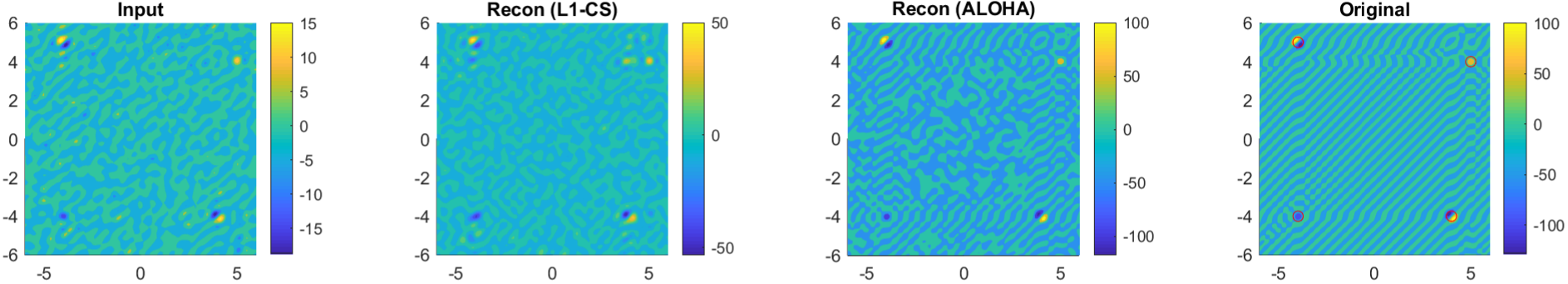}\label{fig2c}}
\\
\caption{Source reconstruction using noisy data with AWGN ($3$dB SNR). Left-to-right: Input (with sub-sampled data without regularization), Recon. L1-CS (with sub-sampled data and L1-CS-based sparse regularization), Recon. ALOHA (with sub-sampled data and ALOHA-based regularization), and Original (with fully sampled data).}
\label{fig2}
\end{center}
\end{figure}

Numerical simulations in Fig. \ref{fig2} show that the conventional algorithm is unable to properly identify point sources with  sub-sampled data. The L1-CS approach improves the overall signal strength and compensates for the losses in terms of PSNR even with $5\%$ of the measurements, but the gain is not effective enough. The performance of the L1-CS method is susceptible to the number of available measurements. The speckle field induced by the sub-sampling has random hot spots with typical diameters comparable to the size of the sources. As a result, it is challenging to distinguish the sources from the hot spots.
The proposed ALOHA-based strategy, on the other hand, effectively reconstructs the sources using sparse measurements and preserves the quality even with just $5\%$ of the original measurements. 

\begin{figure}[tbh!]
\begin{center}
\subfigure[Noise-free data]{\includegraphics[width = 1\columnwidth]{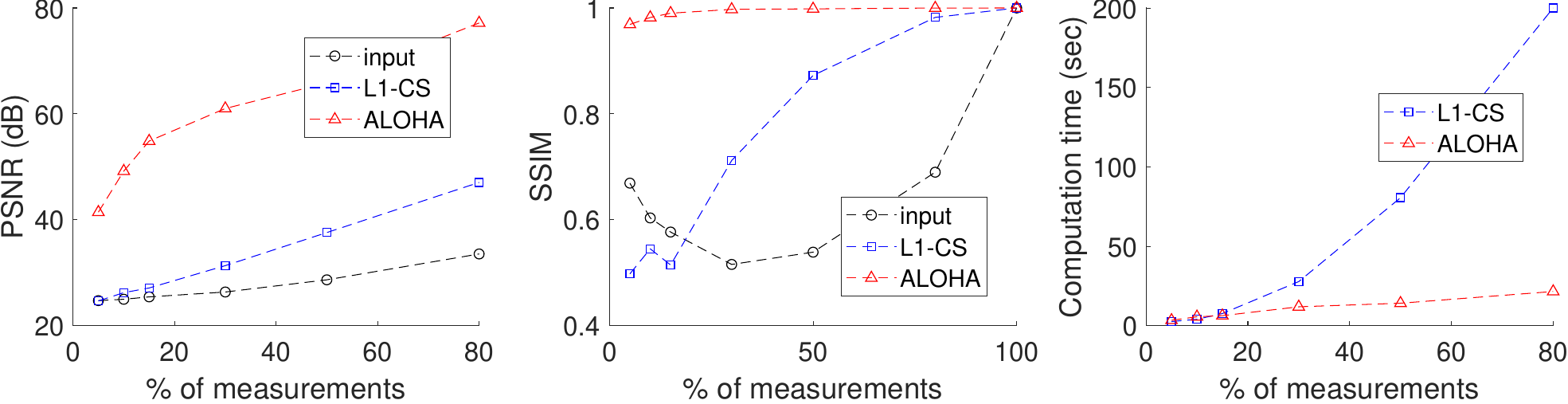}\label{fig3a}}
\\
\subfigure[Noisy data with AWGN ($3$dB SNR)]{\includegraphics[width = 1\columnwidth]{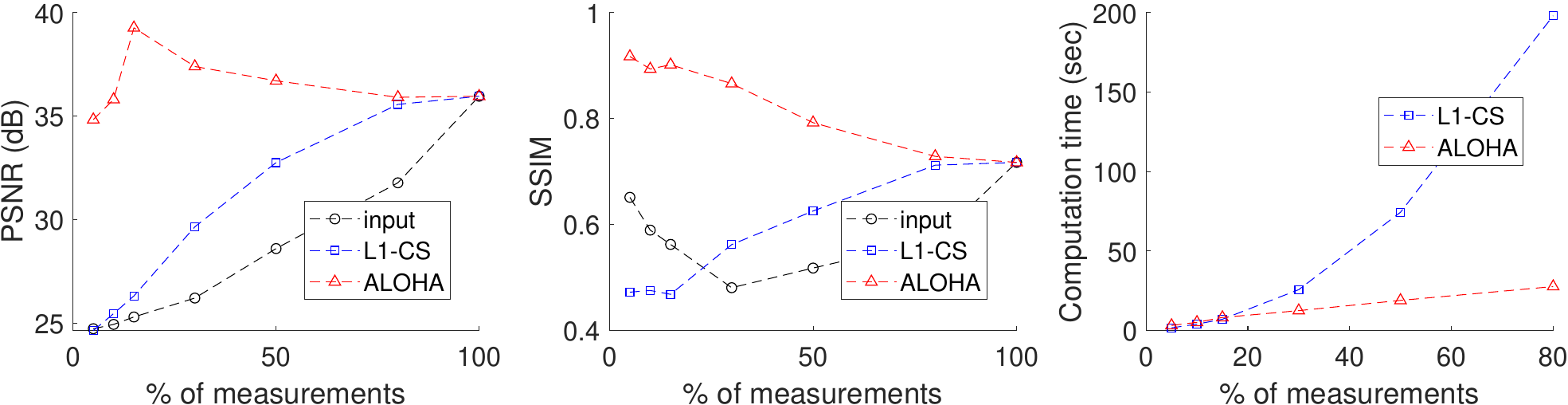}\label{fig3b}}
\\
\subfigure[PSNR and SSIM vs. SNR]{\includegraphics[width = 1\columnwidth]{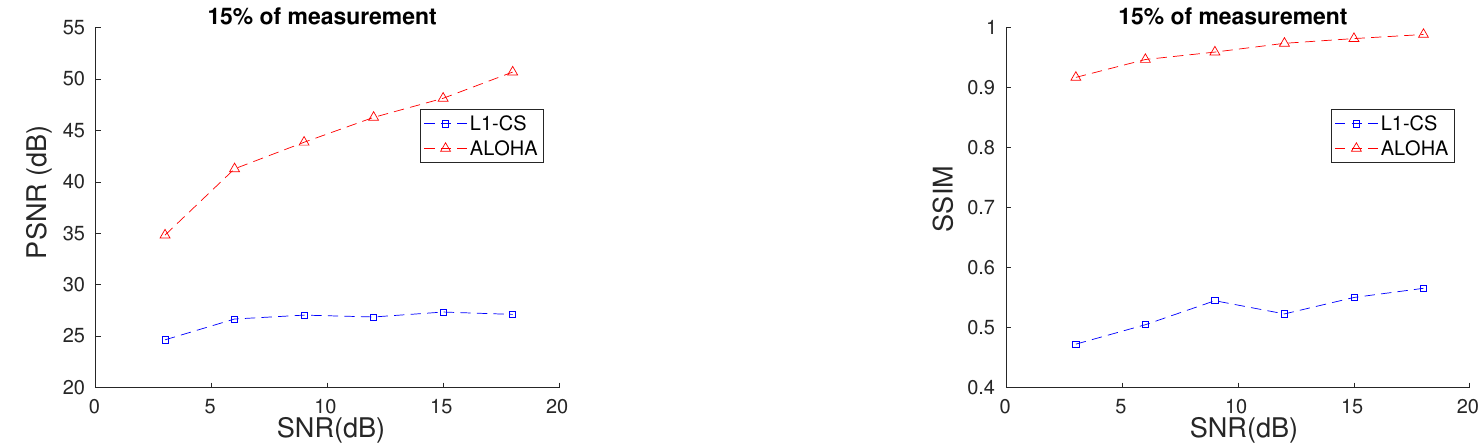}\label{fig3c}}
\caption{Quantitative performance analysis for noise-free and noisy data.}
\label{fig3}
\end{center}
\end{figure}

For further analysis, PSNR, SSIM, and computational time are evaluated for measurements with and without noise. Table \ref{tbl:uni_reg_res} provides a quantitative performance comparison of the sparse reconstruction algorithms for $5\%$, $10\%$, $15\%$, and $30\%$ measurements. Fig. \ref{fig3} substantiates the superiority of the proposed approach for both noise-free (Fig. \ref{fig3a}) and noisy (Figs. \ref{fig3b}-\ref{fig3c}) scenarios. Since the target spectrum only consists of point sources, fewer measurements produce the speckle field in the background medium, which lowers the structural similarity. As a result, for measurements under $50\%$, the structural similarity of the sub-sampled reconstruction decreases as the number of measurements rises, but it begins to rise for measurements beyond $50\%$.
Although the L1-CS method increases PSNR at higher sampling rates, the structural similarity is not improved since its SSIM is smaller than the sub-sampled input in some cases. In contrast, the proposed ALOHA-based approach produces excellent PNSR and SSIM results even with  $5\%$ of measurements. Interestingly, with $100\%$ noisy measurements, all curves meet at the same point (i.e., maximum PSNR and SSIM); however, for sub-sampled noisy measurements, ALOHA not only recovers the lost signal but also lowers the noise (see PSNR and SSIM curves in Fig. \ref{fig3b}). 

\begin{table}[!htb]
    \centering
    \caption{Quantitative performance comparison.}\label{tbl:uni_reg_res}
    \resizebox{1\columnwidth}{!}{
        \begin{tabular}{|c|c|c|c|c|c|c|}
        \hline
        \textbf{Metric} & \textbf{SNR (dB)} & \textbf{Algorithm/Measurements} &
        \textbf{5\%} & \textbf{10\%} & \textbf{15\%} & \textbf{30\%} \\ \hline\hline
        \multirow{6}{*}{PSNR (dB)} & \multirow{3}{*}{$\infty$} &Sub-sampled Input & $24.7407$ & $24.9904$ & $25.4443$ & $26.3654$\\\cline{3-7}
        & &  L1-CS &$24.6796$ & $26.2436$ & $27.1251$ & $31.3453$\\\cline{3-7}
        & &  ALOHA &$41.4397$ & $49.1918$ & $54.8742$ & $61.0166$\\
        \cline{2-7}
        &\multirow{3}{*}{3} &  Sub-sampled Input &$24.7387$ & $24.9463$ & $25.3056$ & $26.2187$\\\cline{3-7}
        & &  L1-CS &$24.6485$ & $25.4674$ & $26.3238$ & $29.6587$\\\cline{3-7}
         & &  ALOHA &$34.8373$ & $35.8095$ & $39.2726$ & $37.3999$\\\hline\hline
        \multirow{6}{*}{SSIM} & \multirow{3}{*}{$\infty$} &Sub-sampled Input & $0.6692$ & $0.6035$ & $0.5769$ & $0.5157$\\\cline{3-7}
        & &  L1-CS &$0.4979$ & $0.5452$ & $0.5146$ & $0.7120$\\\cline{3-7}
        & &  ALOHA &$0.9689$ & $0.9820$ & $0.9901$ & $0.9974$\\
        \cline{2-7}
        &\multirow{3}{*}{3} &  Sub-sampled Input &$0.6515$ & $0.5896$ & $0.5626$ & $0.4809$\\\cline{3-7}
        & &  L1-CS &$0.4725$ & $0.4757$ & $0.4684$ & $0.5623$\\\cline{3-7}
         & &  ALOHA &$0.9173$ & $0.8937$ & $0.9018$ & $0.8661$\\\hline\hline
        \end{tabular}}
\end{table}

The computational cost of ALOHA is also better than L1-CS in terms of walk-clock run-time  (see Figs. \ref{fig3a}-\ref{fig3b} (right column)). 
The L1-CS method has a computational cost exponentially proportional to the number of measurements, whereas that for ALOHA is linear. This substantiates the robustness of the proposed approach. Finally, PSNR and SSIM versus SNR plots (Fig. \ref{fig3c}) for $15\%$ measurements show that the performance of ALOHA is directly proportional to the SNR.

\section{Conclusions}\label{s:conc}

We propose a two-stage numerical scheme for reconstructing multipolar sources using sparse multi-frequency far-field data. First, the sparse data is enriched using a structured Hankel matrix completion approach. The sources are then reconstructed using a Fourier inversion algorithm from enriched data at the Nyquist rate. The measurement recovery problem is recast as the recovery problem for the missing spectrum of signals with a finite rate of innovations, which is solved using an annihilating filter-based structured Hankel matrix completion approach (ALOHA). Simulation results provide evidence that the proposed algorithm is much superior to sparsity-promoting L1 compressed sensing regularization. To that end, the proposed numerical scheme provides simulation results with a $10.1$dB PSNR ($41\%$) and a $0.266$ SSIM ($40.8\%$) increase in the worst-case scenario with a linear computing cost for both noise-free and extremely noisy data. To that end, the proposed numerical scheme provides simulation results with a $10.1$dB PSNR ($41\%$) and a $0.266$ SSIM ($40.8\%$) increase in the worst-case scenario with a linear computing cost for both noise-free and extremely noisy data.

\section*{Acknowledgment}

The authors are thankful to Prof. Jong Chul Ye and Prof. Kyong Hwan Jin for providing the MATLAB code for ALOHA.


\begin{thebibliography}{10}
\providecommand{\url}[1]{#1}
\csname url@samestyle\endcsname
\providecommand{\newblock}{\relax}
\providecommand{\bibinfo}[2]{#2}
\providecommand{\BIBentrySTDinterwordspacing}{\spaceskip=0pt\relax}
\providecommand{\BIBentryALTinterwordstretchfactor}{4}
\providecommand{\BIBentryALTinterwordspacing}{\spaceskip=\fontdimen2\font plus
\BIBentryALTinterwordstretchfactor\fontdimen3\font minus
  \fontdimen4\font\relax}
\providecommand{\BIBforeignlanguage}[2]{{%
\expandafter\ifx\csname l@#1\endcsname\relax
\typeout{** WARNING: IEEEtran.bst: No hyphenation pattern has been}%
\typeout{** loaded for the language `#1'. Using the pattern for}%
\typeout{** the default language instead.}%
\else
\language=\csname l@#1\endcsname
\fi
#2}}
\providecommand{\BIBdecl}{\relax}
\BIBdecl

\bibitem{Thio23}
B.~J. Thio, A.~S. Aberra, G.~E. Dessert, and W.~M. Grill, ``Ideal current
  dipoles are appropriate source representations for simulating neurons for
  intracranial recordings,'' \emph{Clinical Neurophysiology}, vol. 145, pp.
  26--35, 2023.

\bibitem{Beltrachini21}
L.~Beltrachini, N.~von Ellenrieder, R.~Eichardt, and J.~Haueisen, ``Optimal
  design of on-scalp electromagnetic sensor arrays for brain source
  localisation,'' \emph{Human Brain Mapping}, vol.~42, no.~15, pp. 4869--4879,
  2021.

\bibitem{Ammari12}
H.~Ammari, E.~Bretin, V.~Jugnon, and A.~Wahab, ``Photoacoustic imaging for
  attenuating acoustic media,'' in \emph{Mathematical Modeling in Biomedical
  Imaging II: Optical, Ultrasound, and Opto-Acoustic Tomographies}, H.~Ammari,
  Ed.\hskip 1em plus 0.5em minus 0.4em\relax Berlin, Heidelberg: Springer,
  2012, pp. 57--84.

\bibitem{Takahashi22}
S.~Takahashi, K.~Suzuki, T.~Hanabusa, and S.~Kidera, ``Microwave subsurface
  imaging method by incorporating radar and tomographic approaches,''
  \emph{IEEE Transactions on Antennas and Propagation}, vol.~70, no.~11, pp.
  11\,009--11\,023, 2022.

\bibitem{Leone18}
G.~Leone, M.~A. Maisto, and R.~Pierri, ``Application of inverse source
  reconstruction to conformal antennas synthesis,'' \emph{IEEE Transactions on
  Antennas and Propagation}, vol.~66, no.~3, pp. 1436--1445, 2018.


\bibitem{Badia02}
A.~E. Badia and T.~Ha-Duong, ``On an inverse source problem for the heat
  equation. application to a pollution detection problem,'' \emph{Journal of
  Inverse and Ill-posed Problems}, vol.~10, no.~6, pp. 585--599, 2002.

\bibitem{AmmariN}
H.~Ammari, E.~Bretin, J.~Garnier, and A.~Wahab, ``Noise source localization in
  an attenuating medium,'' \emph{SIAM Journal on Applied Mathematics}, vol.~72,
  no.~1, pp. 317--336, 2012.

\bibitem{Jerbi04}
K.~Jerbi, S.~Baillet, J.~Mosher, G.~Nolte, L.~Garnero, and R.~Leahy,
  ``Localization of realistic cortical activity in {MEG} using current
  multipoles,'' \emph{NeuroImage}, vol.~22, no.~2, pp. 779--793, 2004.

\bibitem{Yoo2018}
J.~Yoo, A.~Wahab, and J.~C. Ye, ``A mathematical framework for deep learning in
  elastic source imaging,'' \emph{SIAM Journal on Applied Mathematics},
  vol.~78, no.~5, pp. 2791--2818, 2018.

\bibitem{Devaney}
A.~J. Devaney, E.~A. Marengo, and M.~Li, ``Inverse source problem in
  nonhomogeneous background media,'' \emph{SIAM Journal on Applied
  Mathematics}, vol.~67, no.~5, pp. 1353--1378, 2007.

\bibitem{Beltrachini19}
L.~Beltrachini, ``A finite element solution of the forward problem in {EEG} for
  multipolar sources,'' \emph{IEEE Transactions on Neural Systems and
  Rehabilitation Engineering}, vol.~27, no.~3, pp. 368--377, 2019.


\bibitem{Griesmaier017b}
R.~Griesmaier and C.~Schmiedecke, ``A multifrequency {MUSIC} algorithm for
  locating small inhomogeneities in inverse scattering,'' \emph{Inverse
  Problems}, vol.~33, no.~3, p. 035015, 2017.

\bibitem{Bleistein}
N.~Bleistein and J.~K. Cohen, ``Nonuniqueness in the inverse source problem in
  acoustics and electromagnetics,'' \emph{Journal of Mathematical Physics},
  vol.~18, no.~2, pp. 194--201, 1977.


\bibitem{Saman}
A.~Wahab, A.~Rasheed, R.~Nawaz, and S.~Anjum, ``Localization of extended
  current source with finite frequencies,'' \emph{Comptes Rendus Mathematique},
  vol. 352, no.~11, pp. 917--921, 2014.

\bibitem{Kusiak1}
S.~Kusiak and J.~Sylvester, ``The convex scattering support in a background
  medium,'' \emph{SIAM Journal on Mathematical Analysis}, vol.~36, no.~4, pp.
  1142--1158, 2005.

\bibitem{Eller09}
M.~Eller and N.~P. Valdivia, ``Acoustic source identification using multiple
  frequency information,'' \emph{Inverse Problems}, vol.~25, no.~11, p. 115005,
  2009.


\bibitem{Bao15SNA}
G.~Bao, S.~Lu, W.~Rundell, and B.~Xu, ``A recursive algorithm for
  multifrequency acoustic inverse source problems,'' \emph{SIAM Journal on
  Numerical Analysis}, vol.~53, no.~3, pp. 1608--1628, 2015.

\bibitem{Alzaalig20}
A.~Alzaalig, G.~Hu, X.~Liu, and J.~Sun, ``Fast acoustic source imaging using
  multi-frequency sparse data,'' \emph{Inverse Problems}, vol.~36, no.~2, p.
  025009, 2020.

\bibitem{Wang}
X.~Wang, Y.~Guo, D.~Zhang, and H.~Liu, ``Fourier method for recovering acoustic
  sources from multi-frequency far-field data,'' \emph{Inverse Problems},
  vol.~33, no.~3, p. 035001, 2017.



\bibitem{Griesmaier17}
R.~Griesmaier and C.~Schmiedecke, ``A factorization method for multifrequency
  inverse source problems with sparse far field measurements,'' \emph{SIAM
  Journal on Imaging Sciences}, vol.~10, no.~4, pp. 2119--2139, 2017.

\bibitem{Potts13}
D.~Potts and M.~Tasche, ``Parameter estimation for nonincreasing exponential
  sums by prony-like methods,'' \emph{Linear Algebra and its Applications},
  vol. 439, no.~4, pp. 1024--1039, 2013, 17th Conference of the International
  Linear Algebra Society, Braunschweig, Germany, August 2011.

\bibitem{Fannjiang10}
A.~C. Fannjiang, T.~Strohmer, and P.~Yan, ``Compressed remote sensing of sparse
  objects,'' \emph{SIAM Journal on Imaging Sciences}, vol.~3, no.~3, pp.
  595--618, 2010.

\bibitem{vetterli2002sampling}
M.~Vetterli, P.~Marziliano, and T.~Blu, ``Sampling signals with finite rate of
  innovation,'' \emph{IEEE Transactions on Signal Processing}, vol.~50, no.~6,
  pp. 1417--1428, 2002.

\bibitem{jin2015annihilating}
K.~H. Jin and J.~C. Ye, ``Annihilating filter-based low-rank {H}ankel matrix
  approach for image inpainting,'' \emph{IEEE Transactions on Image
  Processing}, vol.~24, no.~11, pp. 3498--3511, 2015.

\bibitem{ye2016compressive}
J.~C. Ye, J.~M. Kim, K.~H. Jin, and K.~Lee, ``Compressive sampling using
  annihilating filter-based low-rank interpolation,'' \emph{IEEE Transactions
  on Information Theory}, vol.~63, no.~2, pp. 777--801, 2017.

\bibitem{ADMM}
S.~Boyd, N.~Parikh, E.~Chu, B.~Peleato, and J.~Eckstein, ``Distributed
  optimization and statistical learning via the alternating direction method of
  multipliers,'' \emph{Foundations and Trends in Machine Learning}, vol.~3,
  no.~1, pp. 1--122, 2011.

\bibitem{8187544}
F.~Bach, R.~Jenatton, J.~Mairal, and G.~Obozinski, \emph{Optimization with
  Sparsity-Inducing Penalties}.\hskip 1em plus 0.5em minus 0.4em\relax Now
  Publishers, 2012.

\bibitem{cvx}
M.~Grant and S.~Boyd, ``{CVX}: Matlab software for disciplined convex
  programming, version 2.1,'' \url{http://cvxr.com/cvx}, Mar. 2014.

\end{thebibliography}
\end{document}